\documentstyle[aps,epsf%
,twocolumn%
]{revtex}
\begin{document}
\draft
\title{Self-Organized Criticality with Complex Scaling Exponents in
the Train Model}
\author{Franz-Josef Elmer}
\address{Institut f\"ur Physik, Universit\"at
   Basel, CH-4056 Basel, Switzerland}
\date{September, 1997}
\maketitle
\begin{abstract}
The train model which is a variant of the Burridge-Knopoff earthquake
model is investigated for a velocity-strengthening friction law.  It
shows self-organized criticality with complex scaling exponents. That
is, the probability density function of the avalanche strength is a
power law times a log-periodic function. Exact results (scaling
exponent: $3/2+2\pi i/\ln 4$) are found for a nonlocal cellular
automaton which approximates the overdamped train model. Further the
influence of random static friction is discussed.
\end{abstract}
\pacs{PACS numbers: 64.60.Lx, 05.70.Ln}

\narrowtext

Ten years ago Bak, Tang, and Wiesenfeld showed that a weakly driven
dissipative system with many metastable states can organize itself
into a critical state in the sense of a second-order phase transition
\cite{bak87}. Because criticality is characterized by scale
invariance they expect power-law behavior. For example the
probability density function of the strength $S$ of the
restructuring events (avalanches) should scale like $1/S^B$ where $B$
is some positive real number.

In continuous scale invariance the scaling factor $\lambda$ can be
arbitrary.  This invariance is partially broken if $\lambda$ is
restricted to a specific value $\tilde{\lambda}$ and its integer
powers $\tilde{\lambda}^n$. This {\em discrete scale invariance\/}
has a profound consequence \cite{sor97}. It leads to {\em complex\/}
scaling exponents $B+iC$, where $C=2\pi/\ln\tilde{\lambda}$.  More
precisely: The scaling function has the form $S^{-B}f(C\ln S)$, where
$f$ is a $2\pi$-periodic function.  Sornette and collaborators have
shown that such scaling functions are common in many areas, e.g.
fractals, deterministic chaos, dendritic growth and rupture
\cite{sor97}.

\begin{figure}
\epsfxsize=80mm\epsffile{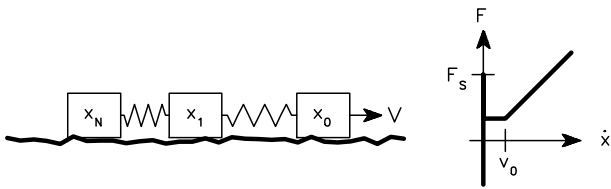}
\vspace{5mm}
\caption[Train model]{\protect\label{f.tmod}The train model and the 
phenomenological friction law (\ref{F}).}
\end{figure}

In this paper we present two models which exhibit self-organized
criticality (SOC) with complex scaling exponents. The first model is
the train model \cite{sou92} which is a variant of the well-known
Burridge-Knopoff (BK) earthquake model \cite{bur67}. In the
literature the BK model and the train model are treated as examples
of weakly driven dissipative systems with many metastable states. In
both models power laws for the avalanche (earthquake) statistics have
been found. In most of these studies an unrealistic purely
velocity-weakening friction law is used. But any phenomenological
friction law has to be velocity-strengthening for large velocities. 
Here we use a {\em realistic\/} friction law (see Fig.~\ref{f.tmod})
which is a velocity independent Coulomb law for small velocities. For
velocities larger than $v_0$ it is proportional to the velocity
\cite{rem0}. Our second model is a nonlocal cellular automaton which
approximates the train model in the overdamped limit. 

The train model is a finite chain of $N+1$ blocks on a rough surface.
The blocks are coupled by springs and the interaction with the
surface is described by a phenomenological dry-friction law $F$ (see
Fig.~\ref{f.tmod}):
\begin{equation}
  M\ddot x_j+F(\dot x_j)=\kappa(x_{j+1}-2x_j+x_{j-1}),\quad
   j=1,\ldots,N,
  \label{T.model}
\end{equation}
where $M$ is the mass of a block, $x_j$ is the position of block $j$,
and $\kappa$ is the stiffness of the springs. The system is driven by
pulling block zero with a very small velocity $v$, i.e., $x_0=vt$.
The other end of the chain is free, i.e., $x_{N+1}=x_N$.
Our friction law $F$ reads:
\begin{equation}
  F(\dot x)=\left\{\begin{array}{ll}
    (-\infty,F_S]&\mbox{if $\dot x=0$};\\
    \gamma v_0&\mbox{if $0<\dot x<v_0$};\\
    \gamma \dot x&\mbox{if $\dot x>v_0$}.\end{array}\right.
  \label{F}
\end{equation} 
A resting block starts sliding if the sum of the spring forces is
larger than $F_S$. The friction law does not allow backward motions
because the static friction can take any negative value. The kinetic
friction force is a monotonically increasing function of the
velocity. We assume $\gamma v_0<F_S$ otherwise the chain would not
show avalanches.
In the simulation we drive the system infinitesimally slowly. This
can be achieved in the following way. During an avalanche $x_0$ is
held constant. After the avalanche, when all blocks are at rest
(i.e., $\dot x_j=0$), we set $x_0=(1+\epsilon)F_S/\kappa-x_2+2x_1$,
with $\epsilon\ll 1$. Thus the force on block number one is
$(1+\epsilon)F_S$ and just exceeds the threshold for sliding. Usually
we have chosen $\epsilon=10^{-4}$.

\begin{figure}
\epsfxsize=80mm\epsffile{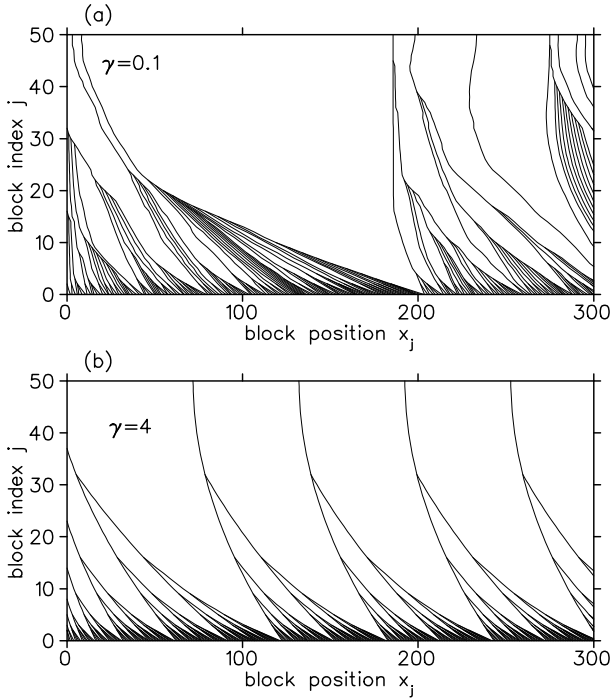}
\vspace{5mm}
\caption[Avalanches]{\protect\label{f.ava}The positions of the blocks
just before an avalanche starts. Several hundreds of avalanches are
shown. The parameters are $N=50$, $v_0=0.01$, $M=F_S=\kappa=1$. 
The initial values are $x_j(0)=0$, for $j=0,\ldots,N$.}
\end{figure}

Figure~\ref{f.ava} shows the evolution of the model for two different
values of the damping constant $\gamma$. The avalanches always start
at the pulling end. They propagate up to a certain block. This is the
reason for the tree-like structures seen in Fig.~\ref{f.ava}. We
characterize the avalanches by two quantities which have a simple
geometric meaning in Fig.~\ref{f.ava}: The length $L$ and the
strength $S$ defined by
\begin{equation}
  S=\sum_{j=1}^N\left|x_j^{\text{after}}-x_j^{\text{before}}\right|,
  \label{S}
\end{equation}
where $x_j^{\text{before}}$ and $x_j^{\text{after}}$ are the position of
the block $j$ before and after the avalanche.  The length $L$ is the
number of blocks that are involved in the avalanche, i.e.,
$x_j^{\text{after}}\neq x_j^{\text{before}}$. For non-system-spanning
avalanches (i.e., $L<N$) this quantity corresponds to the height of
the branching points in the tree. In Fig.~\ref{f.ava} the area below
a branching point is just $S$.

We see a clear distinctive behavior between the underdamped case
($\gamma=0.1$) and the overdamped case ($\gamma=4$)\cite{rem2}. The
underdamped case is characterized by chaotic motion of the chain
leading to an irregular sequence of avalanches even for nonrandom
initial conditions\cite{sou92}. In the overdamped case the motion is
very regular. After the transient which is finished after the first
system-spanning avalanche the same sequence of avalanches reappears
periodically. In both cases SOC with discrete scale invariance
occurs. But the details are different.

\begin{figure}
\epsfxsize=80mm\epsffile{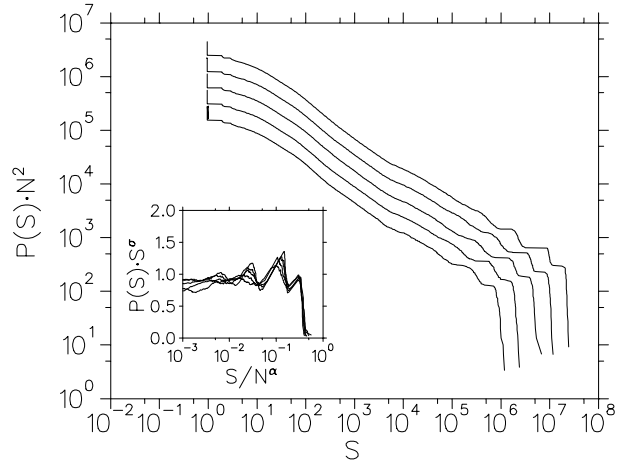}
\vspace{5mm}
\caption[Avalanche statistic]{\protect\label{f.pdf1}The cumulative
density $P(S)$. The parameters are $\gamma=0.1$, $v_0=0.01$,
$M=F_S=\kappa=1$, and $N=530$, 750, 1060, 1500, and 2121. The curves
are shifted by $2\log N$ in order to separate them. The inset shows
the results of finite-size scaling with $\alpha=2.34$ and
$\sigma=1-1/\alpha=0.573$.}
\end{figure}

First we investigate the underdamped case. Fig.~\ref{f.pdf1} shows
the cumulative density $P(S)$ for five different values of the system
length $N$. The cumulative density $P(S)$ is the probability to find
an avalanche that is stronger than $S$. The cumulative densities for
$S$ show steps for large avalanches. Let us assume that $P(S)$
fulfill a usual finite-size scaling ansatz, i.e.,
\begin{equation}
  P(S)=S^{-\sigma}G(S/N^\alpha).
  \label{sf}
\end{equation}
Note that $B=\sigma+1$ because the probability density function is
the derivative of $P$.  The inset of Fig.~\ref{f.pdf1} shows the best
approach to such an ansatz. The value of $\alpha$ was obtained from a
fit of the averaged values of $S$ for the system-spanning avalanches
which are responsible for the last step in the cumulative density.
This fit yields $\alpha=2.34\pm 0.02$. For the value of $\sigma$ we
have chosen $\sigma=1-1/\alpha$ because a sum rule $\langle
S\rangle\sim N$ has to be fulfilled \cite{sou92}.

Although the finite-size scaling ansatz is not completely
satisfactory, one has the impression that for increasing $N$ the
scaling functions are approaching a sawtooth function. In other
words, the cumulative density has steps which becomes steeper and
steeper for increasing $N$ and larger $S$.  The scaling function
shows oscillations which are periodic in the logarithm of the
argument.  This is a clear sign of discrete scale invariance. The
numerically obtained scaling factor is $\tilde{\lambda}=4.3\pm 0.4$.
Thus the complex exponent of the probability density function reads
$B+iC=\sigma+1+2\pi i/\ln\tilde{\lambda}\approx 1.573+4.3i$.

The selfsimilarity of the avalanche tree in Fig.~\ref{f.ava}(b)
reflects the discrete scale invariance in the overdamped case. 
Between two system-spanning avalanches, $2^n-1$ avalanches occur.
Here $n=1+\mbox{int}\,(\log N/\log 2)$, where $\mbox{int}\,(x)$
denotes the largest integer smaller than $x$. There are only $n$
different avalanche lengths and sizes. The $m$-th type of avalanche
occurs $2^{n-m}$ times. Its length is $L=2^{m-1}$.  Thus, the scaling
factor is two. The cumulative density $P(L)$ is in a log-log plot a
staircase with stairs of equal heights and widths \cite{rem3}. A
similar staircase is found for $P(S)$. Here the scaling factor is
$\tilde{\lambda}=2^2=4$ because $S$ corresponds to an area in
Fig.~\ref{f.ava}.  The critical exponents are $\alpha=2$ and
$\sigma=0.5$. Thus $B+iC=3/2+2\pi i/\ln 4$.

The overdamped behavior of the train model can be mimicked by the
following {\em nonlocal\/} cellular automaton. The state of each cell
is given by the block positions $x_j$ and a boolean variable $s_j$
which is $true$ when the block slides, otherwise it is $false$. 
The driving rule is the same as in the train model,
i.e., $x_0=(1+\epsilon)F_S/\kappa-x_2+2x_1$. 

\begin{figure}
\epsfxsize=80mm\epsffile{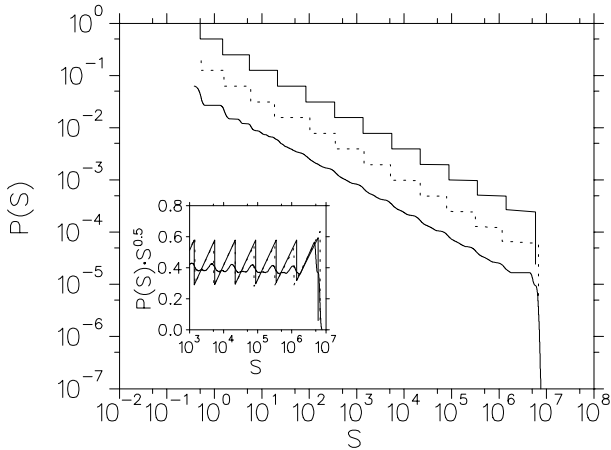}
\vspace{5mm}
\caption[Avalanche statistic]{\protect\label{f.pdf2}The cumulative
density $P(S)$ for the deterministic cellular
automaton for the overdamped train model with and without randomly 
chosen constant static forces $F_S$. The parameters are $\langle
F_S\rangle=\kappa=1$ and $N=2121$.  The curves are shifted in order
to distinguish them. The upper solid curve is the nonrandom case,
i.e. $\Delta F_S=0$. The middle dashed curve is one realization of
the random static friction taken from a Gaussian distribution with
$\Delta F_S=0.1$. The lower solid curve is an average over 1000
realizations of the random static friction.  The inset shows the
scaling functions.}
\end{figure}

The relaxation rules are the following: (i) First the forces
$f_j\equiv \kappa(x_{j-1}-2x_j+x_{j+1}) -F_S$ are calculated. The
variable $s_j$ is {\em true\/} if and only if $f_j>0$. (ii) In the
second step the positions of all sliding blocks are updated 
simultaneously in the following way: 
\begin{equation}
  x_j=x_{j_1}+\frac{x_{j_2}-x_{j_1}}{j_2-j_1}(j-j_1),\quad\mbox{for}\quad
   j_1<j<j_2,
  \label{update.x}
\end{equation}
where $j_1$ and $j_2$ are the left-nearest and right-nearest 
non-sliding blocks. That is, $s_{j_1}=s_{j_2}=false$ and $s_j=true$,
for $j_1<j<j_2$. The consequence of this rule is that
$x_{j-1}+x_{j+1}-2x_j=0$, for $j_1<j<j_2$.  Note that $s_0\equiv
s_{N+1}\equiv false$. If $j_2=N+1$ then $x_{j_2}=x_{j_1}$ in
(\ref{update.x}).  This rule gives the result of a relaxation of
sliding blocks governed by (\ref{T.model}) assuming $v_0=0$ in the
friction law (\ref{F}) and immobile non-sliding blocks.  The slightly
curved lines in Fig~\ref{f.ava}(b) are the effect of $v_0\neq 0$. 
(iii) In the third step the boolean variable $s_j$ is recalculated: A
sliding block still slides and nonsliding block starts to slide for
the same reason as in the first step. That is,
\begin{eqnarray}
 s_j^{\rm new}=s_j^{\rm old}&\vee&(x_{j-1}-2x_j+x_{j+1}-F_S/\kappa>0),
   \nonumber\\&&\mbox{\hspace{30mm}for}\quad j=1,\ldots,N,
  \label{update.s}
\end{eqnarray}
where $\vee$ denotes the boolean operator for inclusive or.  (iv)
Repeat steps (ii) and (iii) until no new block starts to slide in
step (iii). Note that rule (ii) is a {\em nonlocal\/} rule because
$j_1$ and $j_2$ can be arbitrary far from site $j$.  This is in
contrast to most other automata discussed in the field which have
local relaxation rules.

The result of applying these relaxation rules are the following. The
first sliding block is always block number one. In the next cycle
block number two starts to slide. This goes on until no new block
starts to slide. Thus the avalanche length $L$ is the smallest
positive value which fulfills
\begin{equation}
 x_0+(x_{L+1}-x_0)L/(L+1)+x_{L+2}-2x_{L+1}\le F_S/\kappa.  
  \label{Lcon}
\end{equation}
After some straightforward calculations one finds that between two
system-spanning avalanches $2^{n-m}$ avalanches occur. They are
organized in the same binary tree as in the train model [see
Fig.~\ref{f.ava}(b)]. The length and the strength of the $m$-th type
are $L_m=2^{m-1}$ and $S_m=(1+2^{2m-1})F_S/(6\kappa)$, respectively.
Therefore the cellular automaton has exactly the same scaling
exponents and log-periodicity as the overdamped train model.

The behavior of the automaton is very sensitive to nonuniformities in
static friction $F_S$ because the avalanches always stop when
condition (\ref{Lcon}) is just fulfilled. We introduce quenched 
randomness in two different ways leading to two different behaviors. 

In the first way random numbers $F_{Sj}$ from a Gaussian distribution
are assigned to each block. After each system-spanning avalanche the
same sequence of avalanches appears. They are also organized in a
hierarchical manner. The cumulative densities are still stairs (see
Fig~\ref{f.pdf2}) but the heights and the widths of the steps are
fluctuating. Averaging over many realizations of
$\{F_{S1},F_{S2}\ldots,F_{SN}\}$ leads to cumulative densities which
still show log-periodic oscillations (see Fig.~\ref{f.pdf2}). The
oscillation amplitude decreases with the noise level. The phase of
the oscillation changes also but it still does not depend on $N$.

\begin{figure}
\epsfxsize=80mm\epsffile{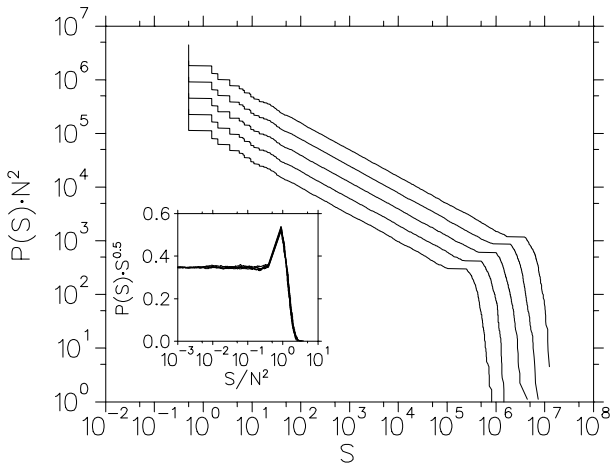}
\vspace{5mm}
\caption[Avalanche statistic]{\protect\label{f.pdf3}The cumulative
density $P(S)$ of the cellular automaton with
randomly chosen static friction $F_S$ which changes after each slide.
The parameters are $\langle F_S\rangle=\kappa=1$, $\Delta
F_c=10^{-3}$, and $N=530$, 750, 1060, 1500, and 2121. The cumulative
densities are shifted by $2\log N$ in order to separate them. The
inset show the scaling function $G$.}
\end{figure}

The automaton shows a different behavior if we assign to each block a
new random number $F_{Sj}$ after a slide.  Figure~\ref{f.pdf3} shows
that the oscillations in the cumulative density vanish completely
even for infinitesimally small noise level\cite{rem4}. Otherwise the
scaling exponents are the same as for the nonrandom case. In the
underdamped case the train model is less sensitive to this kind of
quenched randomness. In simulations of (\ref{T.model}) with $\Delta
F_S/\langle F_S\rangle=10^{-2}$ we still find oscillations but with
smaller amplitudes.

Why have these log-periodicities not be found for purely
velocity-weakening friction laws \cite{sou92}? The main reason may be
that for such laws {\em any\/} sliding motion of the chain or a part
of the chain is {\em unstable\/}.  For a velocity-strengthening
friction force, sound waves with wavelength larger than
$\tilde{L}\equiv 4\pi\sqrt{\kappa M}/\gamma$ are overdamped because
their frequency becomes less than $\gamma/2M$.  Thus avalanches
involving more than $\tilde{L}$ blocks show regular behavior after
some possibly chaotic transients.  This regular behavior is very well
described by the nonlocal cellular automaton. In the overdamped case
(i.e., $\tilde{L}\lesssim 1$) the regular motion successively
amplifies the smallest intrinsic length scale (i.e., $L=1$) by the
factor two. This basic mechanism is responsible for discrete scale
invariance which would be destroyed by the intrinsic instabilities
cause by a velocity-weakening friction force. It is unclear, why in
the underdamped case (i.e., $\tilde{L}\gg 1$) discrete scale
invariance also occur. The mechanism has to be a different one
because the motion of a group of less than $\tilde{L}$ blocks is in
general irregular. Two observations may help to explain this
phenomenon:  (i) The steps in the cumulative density of $L$ do not
occur at powers of two but at $aN/\tilde{\lambda}_L^n$, for
$n=0,\ldots,\tilde{n}$, where $\tilde{\lambda}_L\approx 2$, $a$ is an
$N$-independent number between $1/2$ and $1$, and $\tilde{n}$ is
clearly less than $\ln(aN/\tilde{L})/\ln 2$.  (ii) Although the large
avalanches are organized in a binary tree, their sequence of occurrence
is different.  Let us consider a big avalanche much bigger than
$\tilde{L}$ but not a system-spanning one. Looking for a considerably
bigger avalanche in the past and in the future we find that in the
overdamped case both time intervals are the same whereas in the
underdamped case the time interval in the past is much smaller than
the time interval in the future. In fact only a handful of very tiny
avalanches occur in the time interval in the past.

We have shown that SOC with complex scaling exponents occurs in the
train model with a {\em realistic\/} velocity-strengthening friction
law.  We are confident that other depinning models may also show
discrete scale invariance. The key ingredients are the following: (i)
There should be a bistability between pinning and sliding for the
same local force.  This can be either achieved by inertia of the
pinned objects or by age-dependent pinning (i.e., the pinning force
which increases with the pinning time). (ii) The sliding dynamics
should be nonchaotic with diffusion-like relaxation of
long-wavelength excitations. In BK-like models a
velocity-strengthening friction laws are a necessary condition for
that. (iii) Quenched randomness in the pinning forces should be
absent or weak.  The first two properties are necessary to get
nonlocal deterministic relaxations rules similar to our automaton.
These rules are responsible for discrete scale invariance because
they either amplify successively an intrinsic (microscopic) length
scale up to the system size by a constant factor or vice versa.
Quenched randomness distort these process leading to fluctuations of
the phase of the log-periodicity which may be smeared out if they are
too strong.

\acknowledgments I gratefully acknowledge D. Sornette who introduced
me into the concept of discrete scale invariance, H. Thomas for
critical reading of the manuscript, and the Centro Svizzero di
Calcolo Scientifico at Manno, Switzerland, for doing the simulations
on the NEC SX-3 and SX-4.


\begin{references}
\bibitem{rem2}
There is no clear distinction between underdamped and overdamped
motion because it depends on the wave length of the sound waves. For
$\gamma<\sqrt{\kappa M}\pi/N$ all sound waves are underdamped and for
$\gamma>4\sqrt{\kappa M}$ all sound waves are overdamped.
\bibitem{rem3}
The right most step occurs at $L=N$ which does not fits into this
power-of-two scheme.
\bibitem{rem4}
That is, for $\Delta F_S/\langle F_S\rangle$
approximately below (above) $\epsilon$ the cumulative densities do
(not) oscillate.
\end{references}
\end{document}